\documentclass[aps,prx,preprint,superscriptaddress]{revtex4-1}

\usepackage[english]{babel}
\usepackage{graphicx}
\usepackage{color,xcolor}
\usepackage{dcolumn}
\usepackage{bm}
\usepackage{amsmath,amssymb}
\usepackage{pst-all}
\usepackage{psfrag}
\usepackage{verbatim} 
\usepackage{bm}
\usepackage{natbib}
\usepackage[colorlinks=true,bookmarks=false,citecolor=blue,urlcolor=blue]{hyperref} 
\usepackage{bm}
\usepackage{verbatim} 
\usepackage{morefloats}
\usepackage{graphicx}
\usepackage[mathscr]{euscript}
\usepackage{cleveref}

\def\dfrac{\displaystyle\frac}  
\newcommand{\eps}{\varepsilon}

\renewcommand {\i}{{\rm i}}

\renewcommand {\phi}{\varphi}

\begin{document}

\title{Multipolar third-harmonic generation driven by optically-induced magnetic resonances}

\author{Daria~A.~Smirnova}
\affiliation{Nonlinear Physics Center, Australian National University, Canberra ACT 2601, Australia}

\author{Alexander~B.~Khanikaev}
\affiliation{Department of Physics, Queens College of The City University of New York, Queens, NY 11367, USA}
\affiliation{Department of Physics, The Graduate Center of The City University of New York, NY 10016, USA}

\author{Lev~A.~Smirnov}
\affiliation{Institute of Applied Physics of the Russian Academy of Science, Nizhny Novgorod 603950, Russia}

\author{Yuri~S.~Kivshar}
\email[]{ysk@internode.on.net}
\affiliation{Nonlinear Physics Center, Australian National University, Canberra ACT 2601, Australia}

\begin{abstract}
We analyze third-harmonic generation from high-index dielectric nanoparticles and discuss the basic features and multipolar nature 
of the parametrically generated electromagnetic fields near the Mie-type optical resonances. 
By combining both analytical and numerical methods, we study the nonlinear scattering from simple nanoparticle geometries such 
as spheres and disks in the vicinity of the magnetic dipole resonance. 
We reveal the approaches for manipulating and directing the resonantly enhanced nonlinear emission with subwavelength all-dielectric structures 
that can be of a particular interest for novel designs of nonlinear optical antennas and engineering the magnetic optical nonlinear response at nanoscale.
\end{abstract}


\maketitle


The pursuit for using all-dielectric components as building blocks in nanoscale devices and photonic circuitry
constitutes an important trend in modern nanophotonics~\cite{Jahani2016}. It ultimately aims to circumvent the challenge of Ohmic losses and heating detrimental
to the performance of conventional metal-based plasmonics. Nanostructures made of high-refractive-index semiconductors
and dielectrics exhibit strong interaction with light due to the excitation of the localized Mie-type resonances they sustain~\cite{GarcaEtxarri2011,Schmidt2012}. Shrinking light in high-index and low-loss dielectric nanoparticles,
acting as open optical high-$Q$ resonators (or resonant optical nanoantennas), opens up an access to the optically-induced
response of magnetic nature associated with {\em magnetic} field multipoles. In most previous works on trapped magnetic resonances,
silicon was the primary focus of the material~\cite{Evlyukhin2012,Kuznetsov2012,Staude2013,Zywietz2014,Liu2014,Zywietz2015}.
Remarkably, along with a moderately high refractive index and relatively low absorption at the visible, infrared and telecom frequencies~\cite{Palik},  silicon, both crystalline and amorphous, possesses a strong cubic optical nonlinearity~\cite{Burns1971,Moss1989,Moss1990,Bristow2007,Lin2007,Zhang2007,Ikeda2007,VivienBook,Gai2014}
that makes it suitable for {\em all-dielectric nonlinear nanophotonics}, bringing many intriguing capabilities of the efficient
all-optical light control. Recently, the enhancement of nonlinear response attributed to the magnetic dipole resonances
in silicon nanodisks has been demonstrated experimentally~\cite{Shcherbakov2014,Shcherbakov2015,Shcherbakov2015_2}.

Generally, in the problems of both linear and nonlinear scattering at arbitrary nanoscale objects,
multipole decomposition of the scattered electromagnetic fields (see METHODS) provides a
transparent interpretation for the measurable far-field characteristics, such as radiation efficiency
and radiation patterns, since they are essentially determined by
the interference of dominating excited multipole modes~\cite{Jackson1965,BohrenBook,Dadap1999,Dadap2004,Kujala2008,Petschulat2009,Gonella2011,
Mhlig2011,Huttunen2012,Capretti2012,Grahn2012}. Tuning the contributions of different-order multipole moments is used to engineer the scattering and tailor the emission directionality
of optical nanoantennas~\cite{Liu2012,Rodrigo2013,Staude2013,Fu2013,Poutrina2013,Krasnok2014,Smirnova2014,DavidsonII2015, Dregely2014, Liberal2015}. In particular, the first Kerker condition for overlapped and balanced electric and magnetic dipoles represents an
example for {\em zero backscattering}~\cite{Kerker1983}. Developing this concept, the directionality of the 
scattering can be improved through the interference of properly excited higher-order electric and magnetic modes~\cite{WLiu2014, Hancu2014, Naraghi2015, Alaee2015}. Given the fact that considerable electromagnetic energy can be confined to small volumes in nanostructured materials, nonlinear optical phenomena are of a special interest, and they offer exclusive prospects for engineering fast and strong
optical nonlinearity and controlling light by light. By virtue of subwavelength localization of eigenmodes and resonant character of their excitation, nonlinear effects may become quite pronounced, even at relatively weak external fields~\cite{Kauranen2012}.
Therefore, nonlinear response depends largely on localized resonant effects in nanostructures, allowing for nonlinear optical components to be scaled down in size, which is important for fully functional photonic circuitry. 

\begin{figure}[t!]
\centerline{\mbox{\resizebox{8.2cm}{!}{\includegraphics{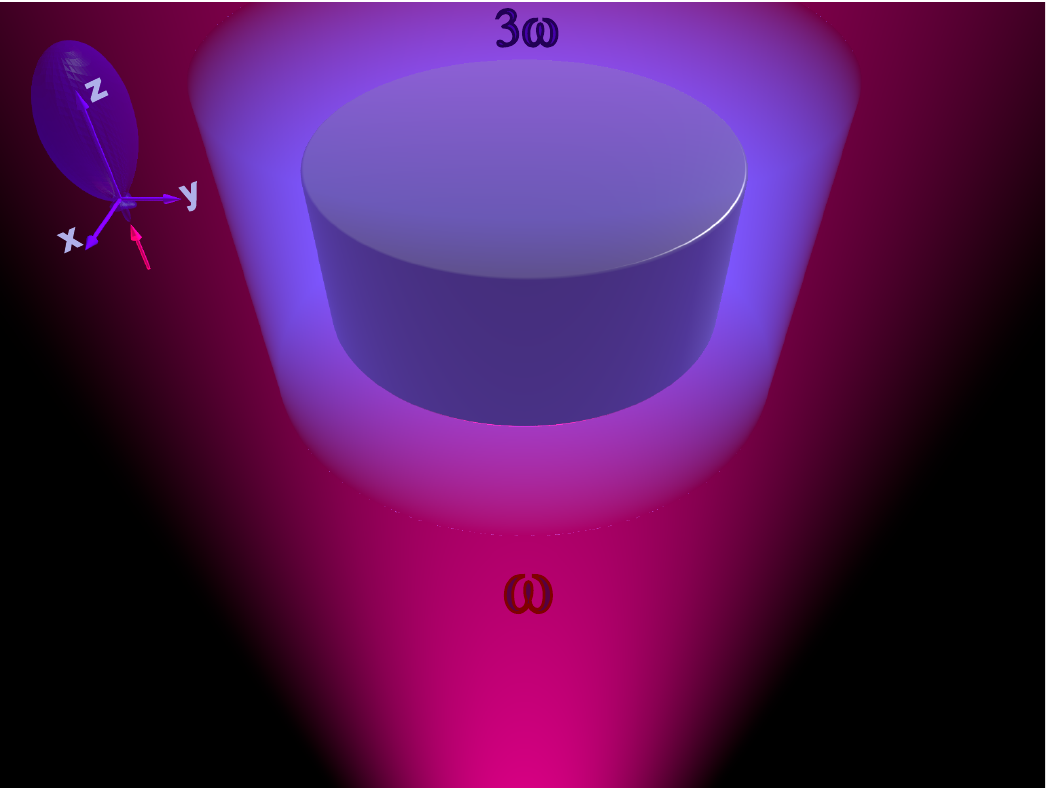}}}}
\caption{{\bf Schematic of the third-harmonic generation from a resonant nanoparticle.}  A dielectric nanodisk is illuminated by laser light from the bottom. The third-harmonic radiation is shown predominantly emitted into the upper half-space, with a forward-directed radiation pattern formed in the far-field, when the magnetic and electric dipolar modes overlap at the fundamental frequency
under the plane wave excitation.
}
\label{fig:fig1}
\end{figure}
%

The efficiency of harmonic generation 
can be strongly enhanced in nanostructures, provided the pump or generated frequency matches the supported resonance~\cite{Boyd}, especially, if the geometry is doubly resonant~\cite{Hayat2007,Banaee2011,Thyagarajan2012,NavarroCia2012,Aouani2012,Ginzburg2012}, i.e. sustains resonances at both the fundamental and harmonic frequencies, and spatial distribution of the nonlinear source is such that it strongly couples the corresponding modes. In this respect, electric dipolar resonance has been widely exploited in deeply subwavelength metallic particles and their composites, 
and thereby plasmonic nanostructures offer a unique playground to study a rich diversity of nonlinear phenomena, including second-harmonic generation (SHG)~\cite{Roke2012,Kauranen2012,Butet2015,Segovia2015}, third-harmonic generation (THG)~\cite{Metzger2012,Hentschel2012,Metzger2014} and four-wave mixing (FWM)~\cite{Zhang2013}. To characterize the nonlinear scattering, the nonlinear Mie theory was developed over recent decades~\cite{Dewitz1996,Pavlyukh2004,deBeer2009,Gonella2011,Butet2012,Roke2012} as an extension of the analytical approach outlined in Refs.~\cite{Dadap1999,Dadap2004} for the Rayleigh limit of SHG from a spherical particle. This theory can be applied to metallic nanoparticles to describe SHG governed by the dominant surface SH polarization source~\cite{Bachelier2010,Wang2009}. 

In the case of metamaterials, specific nonlinear regimes can be achieved due to magnetic optical response of "meta-atoms"~\cite{Shadrivov2015_book,Rose2013,Kruk2015}. Alternatively to lossy metal-based setups constructed to support loop currents, high-index dielectric particles
can present a strong magnetic response and demonstrate magnetic properties associated with Mie resonances. In this way, resonantly enhanced nonlinear effects for nonlinear dielectric particles of sizes comparable with the inner wavelength
are expected around the frequencies of Mie modes.

The main purpose of this paper is twofold. First, we discuss the key theoretical aspects underlying the third-harmonic
generation from high-index dielectric nanoparticles excited near the magnetic dipole resonance (see Fig.~\ref{fig:fig1}) 
by applying the analytical approach and confirming the analytical predictions by the full-scale numerical simulations in the case of simple geometries such as spheres and disks. Second, we reveal the basic mechanisms for manipulating and directing nonlinear scattering with all-dielectric structures that can be of a particular interest for novel designs of nonlinear optical antennas~\cite{Bharadwaj,Novotny2011,Krasnok2013,Krasnok2014_chapter}.

\section*{RESULTS AND DISCUSSIONS}

\subsection*{Third-harmonic generation from a spherical nanoparticle}

To gain physical insight, we start with developing the basic analytical tools of the third-harmonic generation, and consider 
a high-permittivity spherical dielectric particle of radius $a$ and linear refractive index $n$,
placed in homogeneous space and excited by the linearly-polarized plane wave propagating in the $z$ direction, ${\bf E}_i ({\bf r}) = \hat{{\bf x}} E_0 e^{ik_1z}$. 
The interior of the sphere and exterior region are characterized by the 
permittivities $\eps_{2}$ and $\eps_{1}$, and permeabilities $\mu_2$ and $\mu_1$, respectively.
For conceptual clarity, we assume that the materials are nonmagnetic, i.e. $\mu_2 = \mu_1 = \mu_0 $,  
isotropic and lossless. The particle possesses the third-order nonlinearity, bulk and isotropic, and its nonlinear response is described by the nonlinear polarization ${\bf P}^{(3)} = \eps_0 \chi^{(3)}_V {\bf E}^3$, induced by the pump field, where $\chi^{(3)}_V$ is the cubic susceptibility.

The problem of linear light scattering by a sphere is solved using the multipole expansion, well-known as Mie theory~\cite{BohrenBook,Jackson1965}. In accord with the exact Mie solution, the linear scattering spectum features
resonances accompanied by the local field enhancement. If the refractive index is high enough, electric (ED) and magnetic (MD) dipolar resonances are well-separated in frequency and rather narrow 
(we illustrate this case for an example of large $n$ in Fig. S1(a) of Supporting Information), and the electric field profiles within the particle at the resonances are substantially different. The structure of the local fields is known to play an indefeasible role in nonlinear optical effects with both dielectric and plasmonic resonant systems. In the vicinity of resonances, the field distribution inside the particle excited by the plane wave can be approximated by the corresponding specific eigenmode. Here, we focus on the MD resonance, exhibiting a higher quality factor, and elaborate a fully analytical treatment of the nonlinear problem.

\begin{figure}[!t]\centering
 \centerline{\mbox{\resizebox{6.25cm}{!}{\includegraphics{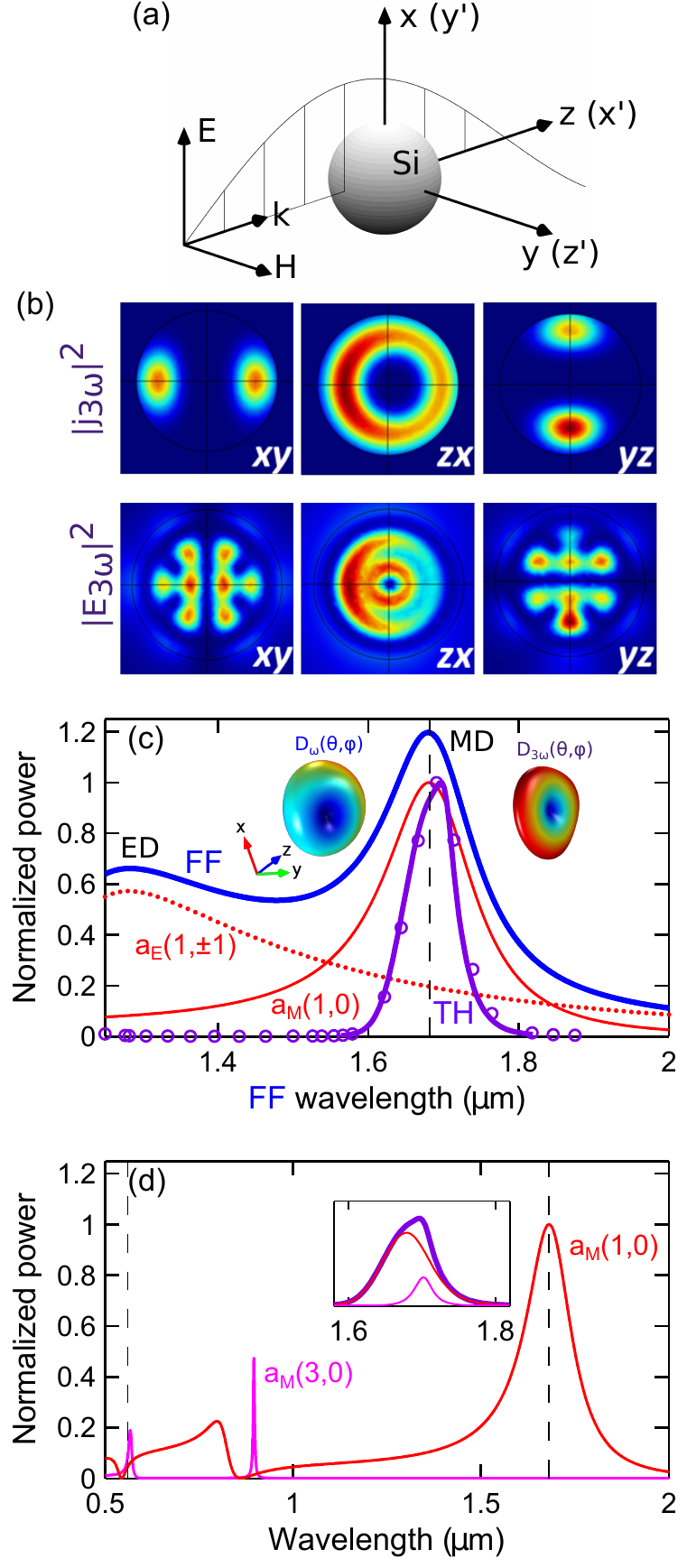}}}}  
  \caption{{\bf Third-harmonic generation by a spherical Si nanoparticle.} (a) Sketch of the geometry. (b) Spatial distributions of the nonlinear current (top) and the intensity of the TH generated field (bottom) at the MD resonance, computed with FEM solver COMSOL for a dielectric particle of radius $a=230$ nm and refractive index $n = 3.5$. (c) Total scattered FF (blue) and radiated TH (purple) powers spectra. Numerical results for THG are displayed with purple circle markers. Red solid and red dotted lines refer to the MD and ED contributions to the FF scattering, respectively. Insets: FF and TH radiation patterns at the MD resonance. (d) Magnetic dipole (red) and magnetic octupole (pink) contibutions to the linear scattering. Inset shows the theoretically obtained magnetic dipole (purple) and magnetic octupole (pink) relative contributions to the TH signal. Powers are normalized to the magnetic dipole maximum values.
}
\label{fig:fig2_2}
\end{figure}

In the single MD mode approximation, the electric field inside the particle at $r<a$ is expressed as
\begin{equation}
\begin{split}
{\bf{E}}^{\text{in}}\left( {\omega} \right) \approx &  {E_0}{ { {A^{\text{in}}_{M}(1,1)j_1(k_2(\omega) r)\left\{ {\bf{X}}_{1,1}( {\theta ,\phi })  +  {\bf{X}}_{1,-1}( {\theta ,\phi }) \right\}
 } } } \\
=  & A(\omega) j_1(k_2(\omega) r)\left\{ \bm {\hat{\theta}} \cos \phi - \sin \phi \cos \theta \bm{\hat{\phi}} \right\},
\end{split}
\label{eq:Field_in_VSH_md}
\end{equation}
where 
$k_2(\omega)=\omega \sqrt{\eps_2(\omega) \mu_2 } = k_0 \sqrt{\eps_{2r} {\mu_{2r}}}$ is the wavenumber inside the sphere, $j_1(k_2(\omega) r)$ is the spherical Bessel function of the first order,
$ {\bf{X}}_{1,1}( {\theta ,\phi }) = \dfrac{1}{4} \sqrt{\dfrac{3}{\pi}} \{ 0, 1, i \cos{\theta} \} \text{exp}(i \phi)$ is the vector spherical harmonic of degree $l=1$ and order $m=1$,
$A(\omega)=E_0 A^{\text{in}}_{M}(1,1) \dfrac{1}{2} \sqrt{\dfrac{3}{\pi}} $, coefficient $A^{\text{in}}_{M}(1,1) $ is known from Mie theory~\cite{BohrenBook,Jackson1965}. For convenience, we rewrite Eq.~\eqref{eq:Field_in_VSH_md} in the spherical coordinate system 
associated with $y$ axis [$z' \leftrightarrow y$, codirected with the magnetic field in the incident plane wave, as shown in Fig.~\ref{fig:fig2_2}(a)]
\begin{equation}
{\bf{E}}^{\text{in}}\left( {\omega} \right) \approx  A(\omega) j_1(k_2(\omega) r) \sin{\theta'} \bm{\hat{\phi'}}.
\end{equation}
In what follows, performing a multipole expansion of the numerically computed fields, 
the multipole moments are calculated in the prime spherical coordinates.

Since the nonlinear medium is isotropic, the nonlinear volume current density induced in the particle at the tripled frequency $3\omega$ 
is 
azimuthal, akin to the electric field,
 $ {\bf j}^{ (3\omega) } = -3i\omega{\bf{P}}^{\left( {3\omega }\right)}$, where ${\bf{P}}^{ (3\omega )} =  \eps_0 \chi^{(3)}_V  (E^{\text{in}}_{\phi'}(\omega))^3 \bm{\hat{\phi'}}$:
\begin{equation}
{\bf j}^{ (3\omega) } = -i 3\omega  \eps_0 \chi^{(3)}_V   A^3(\omega) j^3_1(k_2(\omega) r) \sin^3{\theta'} \bm{\hat{\phi'}}\:.
\end{equation}
Looking for the solution of Maxwell's equations written for the region $r<a$ with this source in the right-hand side
\begin{equation}
\begin{cases}
\nabla \times {\bf E}_{(3\omega)} = 3i\omega \mu_2 {\bf H}_{(3\omega)}, & \\
\nabla \times {\bf H}_{(3\omega)} = -3i\omega \eps_2(3\omega) {\bf E}_{(3\omega)} + {\bf j}^{ (3\omega) }
\end{cases}
\end{equation}
in the form ${\bf E}_{(3\omega)}  = {E}_{(3\omega)} (r, \theta') \bm{\hat{\phi'}}$, we 
obtain the equation
\begin{equation}
\nabla \times \nabla \times {\bf E}_{(3\omega)} - 9 \omega^2 \eps_2(3\omega) \mu_2  {\bf E}_{(3\omega)} = 
{f}(r) \sin^3{\theta'} \bm{\hat{\phi'}} \:,
\end{equation}
where ${f}(r) = 9 \omega^2 \mu_2 \eps_0 \chi^{(3)}_V  A^3(\omega) j^3_1(k_2(\omega) r) $. Accomplishing the vector operations, we get the following scalar equation to solve:
\begin{equation}\label{eq:E3}
\dfrac {\partial^2}{\partial r^2} E_{(3\omega)} +  \dfrac{2}{r} \dfrac {\partial}{\partial r} E_{(3\omega)} +  \dfrac{1}{r^2} \dfrac{\partial}{\partial \theta'} \dfrac{1}{\sin \theta'} \dfrac{\partial}{\partial \theta'}  \sin{\theta'}  E_{(3\omega)}  + 9 \omega^2 \eps_2(3\omega) \mu_2  {E}_{(3\omega)}
= - {f}(r) \sin^3{\theta'}\:.
\end{equation}
Homogeneous Eq.~\eqref{eq:E3} possesses solutions of the form $R_l(3\omega \sqrt{\eps_2(3\omega) \mu_2}r) \dfrac{d \mathcal{Y}_{l0} (\theta')}{d \theta'}$, where the functions $R_l$ and $\mathcal{Y}_{l0}$ satisfy
\begin{equation}
\begin{aligned}
& \dfrac {d^2 R_l}{d r^2}  +  \dfrac{2}{r} \dfrac{d R_l}{d r}  + \left(9 \omega^2 \eps_2(3\omega) \mu_2 -\dfrac{l(l+1)}{r^2}\right) R_l = 0\:,\\
& \dfrac{1}{\sin \theta'} \dfrac{d }{d \theta'} {\sin \theta'}  \dfrac{d \mathcal{Y}_{l0}}{d \theta'}  + l(l+1) \mathcal{Y}_{l0} =0\:,
\end{aligned}
\end{equation}
and the angular functions $\mathcal{Y}_{l0}$ and their derivatives up to $l=3$ are written as
\begin{equation}
\begin{aligned}
& \mathcal{Y}_{10} = \sqrt{\dfrac{3}{4\pi}} \cos{\theta'} \:, \quad \dfrac{d \mathcal{Y}_{10}}{d \theta'} = -\sqrt{\dfrac{3}{4\pi}} \sin{\theta'} ,\:\\
& \mathcal{Y}_{20} = \sqrt{\dfrac{5}{4\pi}} \left(\dfrac{3}{2}\cos^2{\theta'} -\dfrac{1}{2}\right)\:, \quad \dfrac{d \mathcal{Y}_{20}}{d \theta'} =
 -\sqrt{\dfrac{5}{4\pi}} \dfrac{3}{2}\sin{2\theta'} ,\:\\
& \mathcal{Y}_{30} = \sqrt{\dfrac{7}{4\pi}} \left(\dfrac{5}{2}\cos^3{\theta'} -\dfrac{3}{2}\cos{\theta'}\right) \:, \quad
\dfrac{d \mathcal{Y}_{30}}{d \theta'} = \sqrt{\dfrac{7}{4\pi}} \dfrac{15}{2} \left(\sin^3{\theta'} -\dfrac{4}{5}\sin{\theta'}\right)\:.
\end{aligned}
\end{equation}

The angular-dependent part of the source in Eq.~\eqref{eq:E3} can be represented as $\sin^3{\theta'} = \dfrac{2}{15}  \sqrt{\dfrac{4\pi}{7}} \dfrac{d \mathcal{Y}_{30}}{d \theta'} -  \dfrac{4}{5}  \sqrt{\dfrac{4\pi}{3}} \dfrac{d \mathcal{Y}_{10}}{d \theta'} $. Thereby, the solution of Eq.~\eqref{eq:E3} are sought using separation of variables in the form
\begin{equation} \label{eq:E_TH}
E_{(3\omega)} = \mathcal{E}^{(1)}_{3\omega} (r) \dfrac{d \mathcal{Y}_{10}}{d \theta'} +  \mathcal{E}^{(3)}_{3\omega} (r) \dfrac{d \mathcal{Y}_{30}}{d \theta'},
\end{equation}
where two terms correspond to the magnetic dipole (MD) and magnetic octupole (MO) excitations, and for the radial functions we have inhomogeneous equations
\begin{equation}\label{eq:radE}
\begin{aligned}
&  \dfrac {d^2 \mathcal{E}^{(1)}_{3\omega} }{d r^2}  +  \dfrac{2}{r} \dfrac{d \mathcal{E}^{(1)}_{3\omega} }{d r}  + \left(9 \omega^2 \eps_2(3\omega) \mu_2  -\dfrac{2}{r^2}\right) \mathcal{E}^{(1)}_{3\omega}   =    \dfrac{4}{5}  \sqrt{\dfrac{4\pi}{3}} {f}(r) \equiv {f}^{(1)} (r)  ,\:\\
&  \dfrac {d^2 \mathcal{E}^{(3)}_{3\omega} }{d r^2}  +  \dfrac{2}{r} \dfrac{d \mathcal{E}^{(3)}_{3\omega} }{d r}  + \left(9 \omega^2 \eps_2(3\omega) \mu_2  -\dfrac{12}{r^2}\right) \mathcal{E}^{(3)}_{3\omega}   =   -  \dfrac{2}{15}  \sqrt{\dfrac{4\pi}{7}} {f}(r)  \equiv { f}^{(3)} (r) \:.
\end{aligned}
\end{equation}
Matching the fields at the spherical particle surface $r=a$ by imposing boundary conditions of continuity for the tangential field components $E_{{\phi'}(3\omega)}(r,\theta')$ and $H_{{\theta'}(3\omega)}  \propto \dfrac{1}{r} \left[ -\dfrac{\partial }{\partial r}\left(r E_{{\phi'}(3\omega)}\right)\right]$, we can then find relative contributions of the dipole and octupole to the third-harmonic emitted radiation and the 
resultant radiation pattern formed in the far-field.
Following this procedure, we write solutions of Eqs.~\eqref{eq:radE}, denoting $ {K}_{1,2}=3\omega\sqrt{\eps_{1,2}(3\omega) \mu_{1,2} }$, 
\begin{equation}\label{eq:solradE}
\mathcal{E}^{(l=1,3)}_{3\omega} =
\begin{cases}
& C^{(l)}_1 j_l( {K}_{2} r) + 0 \cdot y_l( {K}_{2} r) +
 y_l( {K}_{2} r) \int_{0}^{r} dr'{K}_{2} r'^2 {f}^{(l)}(r') j_l( {K}_{2} r')  \\
& \quad \quad \quad \quad \quad \quad - j_l( {K}_{2} r) \int_{a}^{r} dr'{K}_{2} r'^2 {f}^{(l)}(r') y_l( {K}_{2} r')  , \quad r<a\\
& C^{(l)}_2 h_l^{(1)}( {K}_{1} r), \quad r>a\,
\end{cases}
\end{equation}
From the continuity of the radial functions $\mathcal{E}^{(l=1,3)}_{3\omega} (r)$ and $\dfrac{1}{r}\left[ -\dfrac{\partial }{\partial r}\left(r \mathcal{E}^{(l=1,3)}\right)\right]$ at $r=a$,
we derive the system of equations to calculate coefficients $ C^{(l)}_{1,2}$
\begin{equation}
\begin{cases}
&\!C^{(l)}_1 j_l( {K}_{2} a) + y_l( {K}_{2} a) I^{(l)} \! =  \! C^{(l)}_2 h_l^{(1)}( {K}_{1} a)\:, \\
&\! C^{(l)}_1 \left[a\dfrac{d j_l( {K}_{2} r)}{dr}\Biggr|_{\substack{r=a}} \! +  j_l( {K}_{2} a) \right] + I^{(l)}\left[a\dfrac{d y_l( {K}_{2} r)}{dr}\Biggr|_{\substack{r=a}} \! + y_l( {K}_{2} a)\right] \! \\
& =  \! C^{(l)}_2 \left[a \dfrac{d h_l^{(1)}( {K}_{1} r)}{dr}\Biggr|_{\substack{r=a}}   + h_l^{(1)}( {K}_{1} a) \right]\:,
\end{cases}
\end{equation}
where
\begin{equation}
I^{(l)}= \int_{0}^{a} dr'{K}_{2} r'^2 {f}^{(l)}(r') j_l( {K}_{2} r') \:.
\end{equation}
Particularly, the TH field outside the sphere is a superposition of two outgoing spherical waves 
with the amplitudes ${C^{(1)}_2}$ and ${C^{(3)}_2}$,
\begin{equation}
C^{(l)}_2 =
I^{(l)}  \left[\dfrac{j_l( {K}_{2} a) \dfrac{d y_l( {K}_{2} r)}{dr}\Biggr|_{\substack{r=a}}   -  y_l( {K}_{2} a)  \dfrac{d j_l( {K}_{2} r)}{dr}\Biggr|_{\substack{r=a}}  }
{j_l( {K}_{2} a)  \dfrac{d h_l^{(1)}( {K}_{1} r)}{dr}\Biggr|_{\substack{r=a}}    - h_l^{(1)}( {K}_{1} a)  \dfrac{d j_l( {K}_{2} r)}{dr}\Biggr|_{\substack{r=a}}   } \right]
\:.
\end{equation}
Using the asymptotics $h_l^{(1)}( {K}_{1} r) \sim (-i)^{l+1}\dfrac{e^{iK_1 r}}{K_1 r}$ in Eqs.~\eqref{eq:E_TH},~\eqref{eq:solradE} in the far-field zone allows one to determine 
the directional dependence of the TH radiation $D_{3\omega} (\theta,\varphi)$, as well as the total radiated energy flux.
Remarkably, our analytical description is not restricted to the case of purely real dielectric constants. Within the developed formalism, the linear dielectric permittivity at the third-harmonic frequency is not fixed and can be a complex value with its imaginary part responsible for the linear absorption in dielectric.

The analytically retrieved resonant curve of the TH spectrum depicted in Fig.~\ref{fig:fig2_2}(c) was found to be consistent 
with the results of numerical calculations performed with the help of the finite-element method (FEM) solver COMSOL Multiphysics (see the METHODS section for details) 
for the particle of $n=3.5$.
In our simulations, the homogeneous host medium was described by the relative dielectric permittivity $\eps_{1r} = 1$. 
%
Numerically calculated at the MD resonance nonlinear source and TH generated field distributions are demonstrated in Fig.~\ref{fig:fig2_2}(b). Naturally, the depicted nonlinear current, being essentially the cubed field of the MD mode excited by the pump wave, possesses a notable magnetic dipole moment. As seen in Fig.~\ref{fig:fig2_2}(c), close to the MD resonance results obtained for THG with the use of COMSOL still fit well into our 
analytical model that
may be regarded as a shortcut of the more involved treatment exploiting the Green's function formalism~\cite{Biris2010}. 
From the above analysis, we expect only two multipoles to dominate others in the nonlinear response and visualize the spectral disposition of the generated modes around the tripled frequency.
In Fig.~\ref{fig:fig2_2}(d) we plot contributions from the MD $a_M(1,0)$, $\propto 3 |a_{M}(1,0)|^2/k_0^2 $, and MO $a_M(3,0)$, $\propto 7 |a_{M}(3,0)|^2/k_0^2 $, into the linear scattering over the wide range of wavelengths, based on the exact Mie solution. 
Within the broad MD resonance, both magnetic dipole and magnetic octupole contribute to the THG peak, as shown in the inset of Fig.~\ref{fig:fig2_2}(d). This explains the six-petalled TH near-field structure in Fig.~\ref{fig:fig2_2}(a).

Although in this paper we do not focus specifically on the ED fundamental resonance, it is remarkable that the efficiency of THG at the MD resonance significantly exceeds that at the ED resonance. For a spherical particle, this can be shown analytically, applying the approach sketched above for the MD resonance to the case of the resonant excitation of ED. Still another formal way to quantify this effect is based on the Lorentz reciprocity theorem. Particularly, this method allows one to express the excitation coefficients of spherical multipoles, the radiated field is decomposed into, through the overlap integrals of the nonlinear source and the respective spherical modes. As discussed e.g. in Ref.~\cite{OBrien2015}, such treatment, relating in essence the local-field features and far-field properties, brings in considerable intuition for the optimization of nanostructures to realize strong nonlinear response (emphasizing two key ingredients -- the local field enhancement and modal overlaps). Both the approaches (to be presented in detail elsewhere) give an estimate of more than $50$ time difference between the TH radiated powers at the MD and ED resonances for the parameters of Fig.~\ref{fig:fig2_2} and Fig. S1 of Supporting Information. Being also confirmed by the direct full-vector numerical simulations in COMSOL, this striking distinction stems from dissimilarity in spatial distribution of the respective nonlinear sources and quasipotential character of the nonlinear current induced near the ED resonance. Besides, due to the inherent curl of the source polarization induced, magnetic-type resonances in nanostructures will inevitably cause generation of magnetic multipoles, whose relative contribution into nonlinear response is usually overshadowed by more pronounced nonlinear dipole and quadrupole in metallic setups.

Generally, mirror symmetries of a scatter with respect to the planes $x=0$ and $y=0$ impose constraints on phases 
of the dominant excited multipolar moments. In this way, illumination by the $x$-polarized plane wave implies $ a_E(l,-m) = - a_E(l,m),\: a_M(l,-m) = a_M(l,m),\: \text{if}\: m\: \text{is odd} $, and $a_E(l,m) =  a_M(l,m) =0$, if $m$ is even,
for the FF multipoles of the scattered field defined in the $xyz$ coordinate system. In the $x'y'z'$ coordinate system we use, the aforementioned conditions formally take a more cumbersome form: $a_E(l,-m) = a_E(l,m)$, $a_M(l,-m) = a_M(l,m)$ for odd $l$, and $a_E(l,-m) = - a_E(l,m)$, $a_M(l,-m) = - a_M(l,m)$ for even $l$, where nonzero electric and magnetic multipole coefficients respond to odd or even sum of quantum numbers $(l+m)$, respectively, additionally $a_E(l,0) = 0$. Then these relations are also expected to be fulfilled for the TH multipoles of the generated radiation if the bulk nonlinearity is isotropic.
For instance in Fig.~\ref{fig:fig2_2}(c), mapping the multipole decomposition in the vector spherical harmonics basis onto the Cartesian multipole moments reflects that the coefficients $a_E(1, 1) = a_E(1, - 1)$ are combined to form an $x$-oriented electric dipole excitation, and $a_M(1,0)$ stands for the $y$-aligned magnetic dipole, with the power contributions to the linear scattering proportional to $ \propto 3 (|a_{E}(1,1)|^2 + |a_{E}(1,-1)|^2)/k_0^2 $ and $\propto 3 |a_{M}(1,0)|^2/k_0^2 $, respectively.

\begin{figure}[t!]\centering
  \includegraphics[width=0.9\linewidth] {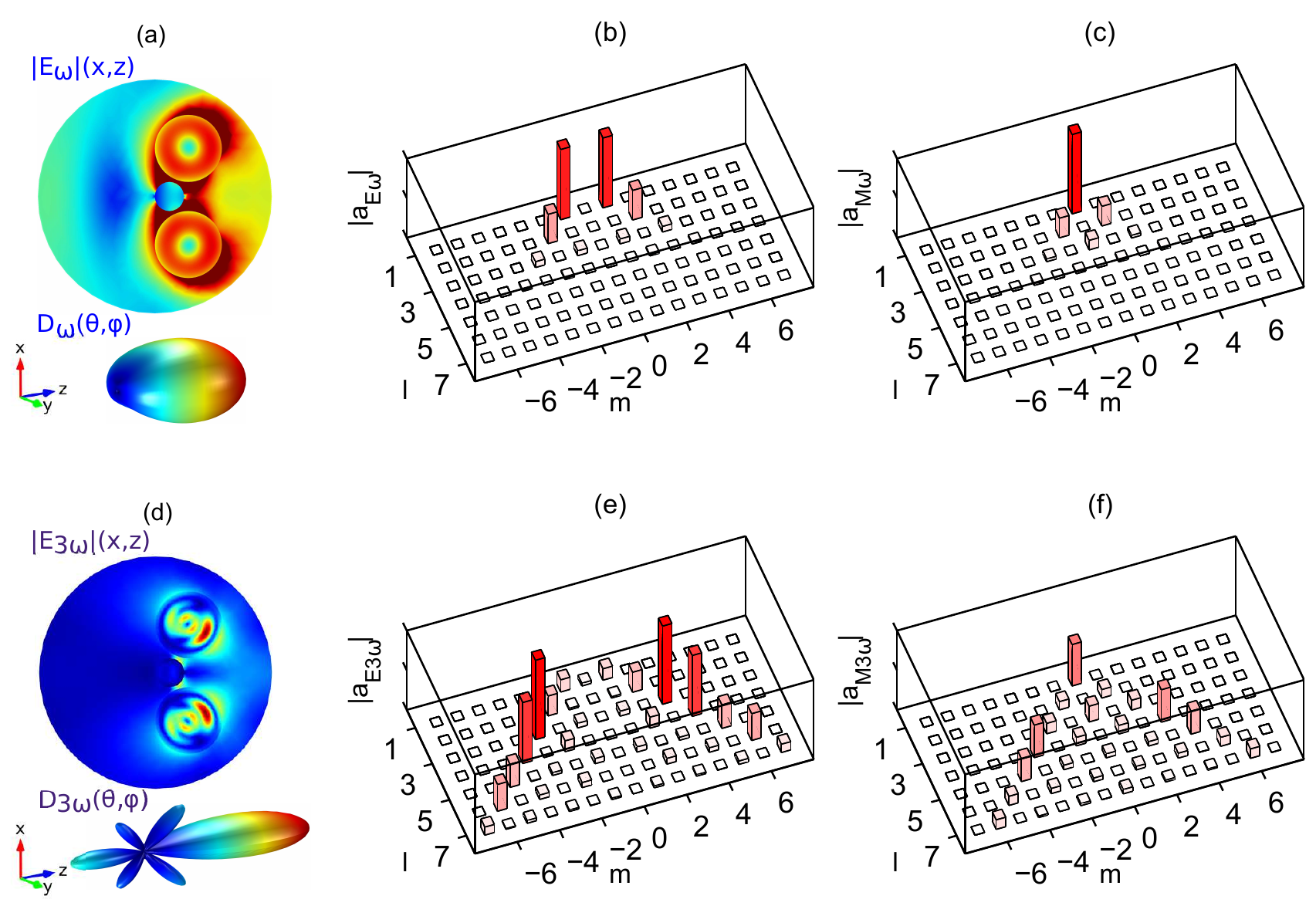}
  \caption{ {\bf Control of linear and nonlinear response with a nanoparticle trimer.} (a,d) Electric field distributions as seen in the $xz$ plane, 3D far-field radiation patterns and magnitudes of the electric (b,e) and (c,f) magnetic multipole moments of the fields at FF (top row) and TH (bottom row) calculated for a trimer of dielectric spheres aligned on their left and excited by the $x$-polarized plane wave incident from the left with the wavelength $1.68$ $\mu$m corresponding to the fundamental MD resonance of a single large sphere.
The radii of the larger spheres and small sphere are $230$ nm and $100$ nm, respectively. Center-to-center distances between
the particles are $660$ nm and $355$ nm, refractive index of the particles $n=3.5$.}
	\label{fig:fig3}
\end{figure}

Note that within our model we assumed the cubic volume nonlinearity of a dielectric scatterer isotropic and characterized by a scalar third-order susceptibility, which is true for amorphous silicon (a-Si). Actual material dispersion and dissipative losses in a-Si, that can be extracted experimentally from ellipsometric measurements, may become nonnegligible at TH frequency and will affect the TH emission spectra. For the parameters of Fig.~\ref{fig:fig2_2}, these dispersive factors only slightly deviate the linear scattering properties at FF, while at TH they influence mainly the region near the ED resonance, where the conversion efficiency is much lower than that near MD resonance. 
However, depending on the optical properties of the sample, they may 
cause a spectral separation of the MD and MO contributions in the nonlinear scattering (similar to that shown in Fig. 1S of Supporting Information) and 
and transformations of TH radiation patterns due to alterations in phases of the generated field multipoles, 
decreasing the total radiated power by a factor of $40$ at the incident intensity $1$ GW/cm$^2$, 
as it follows from the numerical simulations rectified 
with the wavelength-dependent complex-value refractive index of a-Si.

\subsection*{Control over third-harmonic directionality}

For a single spherical particle, we found the radiation pattern of the most efficient THG at the magnetic dipole resonance to be predominantly axially symmetric that does not allow for the emission directionality. Apparently, it also holds for a dimer of spheres under the longitudinal illumination (when the electric field of the incident plane wave is directed along the dimer axis). To shape the TH radiation pattern, one can play with geometry. For instance, placing a small particle near the dimer drastically modifies both the linear and nonlinear response due to inter-particle interaction, as exemplified in Fig.~\ref{fig:fig3}.
The fundamental frequency of the impinging wave is chosen to match the MD resonance of an individual large sphere.
The small particle, acting as an electric dipole at this wavelength when isolated, is magnificantly excited by the resonant co-circulating
electric near-fields of the larger neighbors. As a result, the linear scattering contains both electric and magnetic dipoles and forward-directed, while the TH emission acquires directivity due to the excitation of higher-order multipoles, with six lobes indicating the leading electric octupole $a_E(3,\pm 3)$.

\begin{figure}[t!]
\centerline{\mbox{\resizebox{6.7cm}{!}{\includegraphics{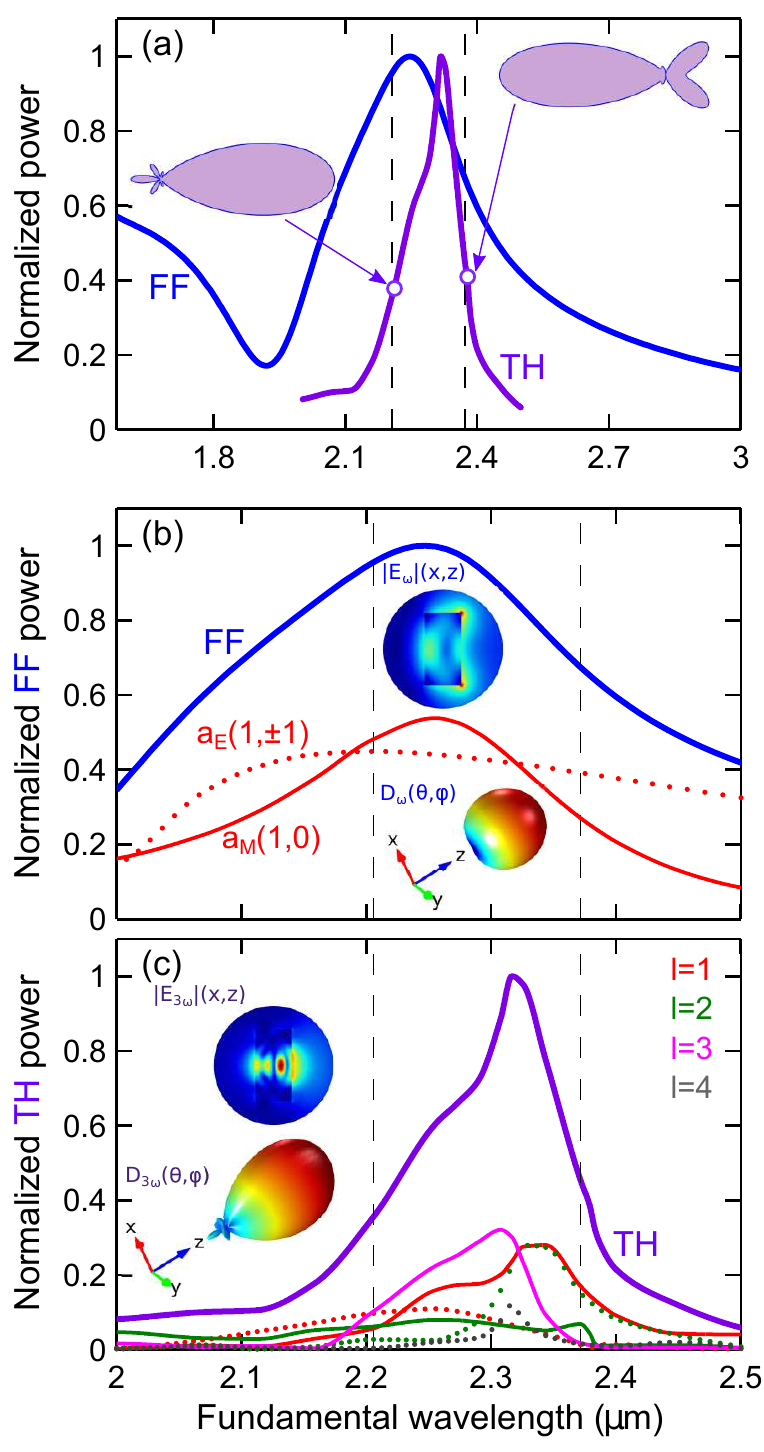}}}}
\caption{ {\bf Resonant nonlinear directional scattering from a Si nanodisk.} (a) Numerical spectra of the total FF scattered (blue line) and TH radiated (purple line) powers calculated for a single disk with radius $r_d = 390$ nm, thickness $h=390$ nm,
and refractive index $n=3.5$ suspended in air. Callouts depict the forward- and backward-directed far-field radiation patterns on the different sides of the TH peak, as indicated by points and thin vertical dashed lines.
(b) Linear scattering spectrum (blue solid curve) overlaid with MD (red solid) and ED (red dotted) contributions to the scattering.
(c) Total TH radiated power (solid purple line) decomposed into multipolar contributions of different orbital numbers $l$. The color solid and dotted curves refer to the main magnetic and electric contributions to the radiation, respectively.
Insets in panels (b) and (c) show 3D radiation patterns and the electric field magnitude distributions in the $xz$ plane at FF and TH at the spectral point corresponding to the forward-directed TH emission, illustrated in Fig.~\ref{fig:fig1}.
Powers are normalized to the magnetic dipole maximum values.}
\label{fig:fig4}
\end{figure}
%

Alternatively, magnetic and electric dipolar eigenmodes can be spectrally overlapped using only one structural element, such as ellipsoid or cylinder, by adjusting aspect-ratio. We calculate numerically the nonlinear response of the disk, presumably made of crystalline Si (c-Si), with the radius and height equal to each other, $r_d = h = 390$ nm. Geometrical parameters of the disk are chosen to escape dramatic influence of dispersion and losses at both fundamental and TH frequencies, and we approximately set the refractive index constant $n=3.5$ in our simulations.
In c-Si the nonlinear polarization at the third-harmonic frequency is given by~\cite{Moss1989}
$P^{(3)}_i = \eps_0\left\{ 3 \chi^{(3)}_{1122} E_i ({\bf E} \cdot {\bf E}) + \left[ \chi^{(3)}_{1111} - 3 \chi^{(3)}_{1122}  \right] E_i^3 \right\}$, with the second term responsible for anisotropy which is a complex-valued frequency-dependent function in general case.
However, for low enough photon energies $\hbar \omega \lesssim  0.6 $ eV ($\lambda \gtrsim 2$ $\mu$m), it was established that two components of the third-order susceptibility are related as $\chi^{(3)}_{1111} = 2.36 \chi^{(3)}_{1122} $~\cite{VivienBook}.
In fact, this anisotropy in the wavelength window considered was not found to result in very substantial alterations
in THG, as compared to the case when the isotropic cubic susceptibility is attributed to the nonlinear disk-shaped
scatterer. The results obtained are summarized in Fig.~\ref{fig:fig4}.

In the nonlinear response, on calculating multipole decomposition, we observe strong excitation of
the following multipole moments up to $l=4$: $a_E(1,\pm 1)$, $a_M(1,0)$, $a_E(2,\pm 2)$, $a_M(2,\pm 1)$, $a_E(3,\pm 3)$, $a_M(3,\pm 2)$, $a_M(3,\pm 0)$, $a_E(4,\pm 4)$, $a_E(4,\pm 2)$, $a_M(4,\pm 3)$, $a_M(4,\pm 1)$. Their amplitudes are sharply frequency-dependent. All others give negligible contributions. Near the intersection point of red dotted and red solid lines in Fig.~\ref{fig:fig4}(b)
[where the strengths of the excited electric and magnetic dipoles are matched (i.e. $p_x \approx \sqrt{\eps_0 \mu_0} m_y$), 
thus producing directional scattering], interference between the nonlinear multipoles of opposite parities gives rise to an unidirectional lobe at TH. On the opposite slope of the TH peak, the lobe switches its direction.

To quantify the efficiency of nonlinear conversion, we define the THG efficiency for a single disk ${\eta}^{\text{THG}}$ as the ratio of the total TH radiated power to the energy flux of the fundamental wave through the physical area of the disk,
${\eta}^{\text{THG}} = \dfrac{W_s^{\text{TH}}}{I_0 \pi r_d^2}$, where $I_0 = \dfrac{|E_0|^2}{2\eta}$ is the incident field intensity.
Near the resonance, the disk exhibits ${\eta}^{\text{THG}} \sim 10^{-5}$ at $I_0 = 1$ GW/cm$^2$ and the value of the nonlinear susceptibility in this frequency range estimated as $ 3\chi^{(3)}_{1122} \approx 2.2 \times 10^{-19}$ (m/V)$^2$, based on the data in Ref.~\cite{Moss1990}. 

Importantly, the efficiency of nonlinear optical processes in Si is known to be limited by two-photon absorption (TPA)~\cite{Lin2007,Shcherbakov2015_2}. At high intensities of light, TPA and subsequent free-carrier absorption (FCA) introduce considerable undesired optical loss and change the refractive index. Generally, TPA is strong in near-infrared, being essentially wavelength-dependent and anisotropic, but drops significantly with increasing the operating wavelength and vanishes at 2.2 $\mu$m. The parameters of Fig.~\ref{fig:fig4} match the range with reduced TPA. However, depending on the operating conditions, the model described in this paper should be extended to account for the relevant effects. In particular, the dependence of the refractive index on the incident field intensity, the so-called self-action associated with Kerr effect and TPA, may shift and broaden Mie resonances and thereby affect THG, raising the need for a more-involved self-consistent model analogous to that developed in Ref.~\cite{Ginzburg2015} for harmonic generation in metallic nanostructures. Nonetheless, for the parameters of Fig.~\ref{fig:fig4}, at the incident energy flux density of $\sim 1$ GW/cm$^2$ the influence of the Kerr effect and TPA on THG can be neglected, since the nonlinear shift of the resonance $\Delta \lambda_0^{\text{(NL)}}$ far exceeds its nonlinear broadening and three orders of magnitude smaller than the width of the Mie resonance (using the data from Ref.~\cite{Lin2007}, the estimate gives $\Delta \lambda_0^{\text{(NL)}} / \lambda_{0} \sim \Delta n^{\text{(NL)}} / n  \sim 10^{-4}$, where $\Delta n^{\text{(NL)}} $ is a nonlinear correction to the refractive index). 

\section*{CONCLUSION}

We have discussed the third-harmonic generation in all-dielectric resonant nanostructures exhibiting optically-induced Mie-type
magnetic response and have demonstrated that such structures offer important tools for tailoring the efficiency and directionality of nonlinear emission at the nanoscale, in particular, by employing the properties of higher-order magnetic multipole moments of the generated field. This framework suggests many novel opportunities for a design of nonlinear subwavelength light sources with reconfigurable radiation characteristics and engineering magnetic optical nonlinearities in nanostructured dielectric metadevices. 

\section*{METHODS} \label{sec:Methods}

{\bf Multipole decomposition.} We adopt the following well-established standard form of the multipole expansion for the scattered (or radiated) field written in the spherical coordinates~\cite{Jackson1965,Grahn2012} (accepting harmonic time-dependence $\sim e^{-i\omega t}$):
\begin{eqnarray}
{\bf E}_{s}\left( {\omega;r,\theta ,\varphi } \right) = E_0\sum_{l=1}^{\infty}\sum_{m=-l}^l i^l\sqrt{\pi (2l+1)}
\Big\{ a_{{M}}(l,m)h_l^{(1)}(kr){\bf X}_{lm}(\theta,\varphi)   \nonumber \\
+ \frac{1}{k}a_{{E}}(l,m)\nabla\times \big[h_l^{(1)}(kr){\bf X}_{lm}(\theta,\varphi)\big]\Big\}, \\
{\bf H}_{{s}}\left( {\omega;r,\theta ,\varphi } \right) = \frac{E_0}{\eta}\sum_{l=1}^{\infty}\sum_{m=-l}^l i^{l-1}\sqrt{\pi (2l+1)}
\Big\{ a_{{E}}(l,m) h_l^{(1)}(kr){\bf X}_{lm}(\theta,\varphi)  \nonumber \\
+\frac{1}{k}a_{{M}}(l,m)\nabla\times\big[h_l^{(1)}(kr){\bf X}_{lm}(\theta,\varphi)\big] \Big\},
\end{eqnarray}
where ${\bf{X}}_{l,m} = \dfrac{-1}{\sqrt{l(l+1)}} \biggl\{ 0, \dfrac{mY_{l,m}(\theta, \phi)}{\sin{\theta}},i\dfrac{\partial Y_{l,m}(\theta, \phi)}{\partial \theta} \biggr\}$ are the vector spherical harmonics expressed through the scalar spherical harmonics $Y_{l,m}(\theta, \phi)$, $h_l^{(1)}$ are the spherical Hankel functions of the first kind with the asymptotics of outgoing spherical waves, $k=\omega\sqrt{\eps\mu} \equiv k_0 \sqrt{\eps_{r}\mu_{r}}$ is the wavenumber in the medium, $\eta = \sqrt{\mu/\eps}$ is the impedance,
$E_0$ is an electric field amplitude factor.
The electric $a_{{E}}(l,m)$ and magnetic $a_{{M}}(l,m)$ multipole cofficients can be found 
either through the volume integration of the source current density distribution, or by using angular integrals of radial components of the numerically pre-calculated scattered field over a spherical surface of some radius $r$ enclosing an isolated scatterer as follows:
\begin{subequations}
\label{eq:aEaM}
\begin{align}
a_{{E}}(l,m) &= \dfrac{(-i)^{l+1}kr}{h_l^{(1)}(kr)E_0\sqrt{\pi(2l+1)l(l+1)}} \displaystyle{\int_0^{2\pi}\int_0^\pi Y^*_{lm}(\theta,\varphi) \hat{{\bf r}}\cdot {\bf E}_{{s}}({\bf r})\sin\theta d\theta d\varphi}, \\
a_{{M}}(l,m) &= \dfrac{(-i)^l\eta kr}{h_l^{(1)}(kr)E_0\sqrt{\pi(2l+1)l(l+1)}}\displaystyle{\int_0^{2\pi}\int_0^{\pi} Y^*_{lm}(\theta,\varphi) \hat{{\bf r}}\cdot{\bf H}_{{s}}({\bf r})\sin\theta d\theta d\varphi}.
\end{align}
\end{subequations}
In terms of these scattering coefficients, the total time-averaged scattered power (energy flow) is given as
\begin{equation}\label{Wsca}
W_s =\dfrac{\pi |E_0|^2}{2\eta k^2} \sum\limits_{l=1}^{\infty} \sum\limits_{m=-l}^{l} {(2l+1) \left( |a_E(l,m)|^2 + |a_M(l,m)|^2 \right)},
\end{equation}
revealing the input of each multipolar excitation.


{\bf Numerical simulation method.} The nonlinear response of nanostructures is simulated in FEM solver COMSOL Multiphysics using two coupled electromagnetic models~\cite{Smirnova2015}, assuming the undepleted pump field. First, we simulate the linear scattering through the excitation of the nanostructure with a plane wave and compute the nonlinear polarization induced in the bulk of the nanostructure. Second, this polarization is employed as a source for the next electromagnetic simulation at the TH frequency, and the generated field is recovered. This method can be applied to nanostructures of complex three-dimensional geometries, and it can be straightforwardly extended to account for the surface nonlinear response.

\section*{Acknowledgements}

The authors thank M. Kauranen, S.H. Mousavi, A.E. Miroshnichenko, and D. Neshev for many useful discussions.
This work was supported by the Australian Research Council. L.A.S. acknowledges support from RFBR, Grants No. 16-32-00635, 16-02-00547.

\bibliographystyle{apsrev4-1}

%

\newpage 
\section*{Supporting Information}
\subsection*{Single-mode approximation}

The analytical theory developed in the paper describes the third-harmonic generation (THG) from a high-permittivity dielectric sphere under the resonant excitation by a plane wave, assuming the nonlinear conversion process is predominantly governed by the corresponding resonant multipolar mode. Generally speaking, such approximation is valid only if the refractive index of the dielectric is sufficiently large and multipolar resonances do not overlap strongly. 
We illustrate this case for the refractive index $n=6$ in Fig. S1. The value of $n$, unpractical in the infrared and optical frequency ranges, is considered here merely for methodological clarity and also for the verification of our analytical formulas. All calculations in the main text are performed with the focus on silicon, which is one of the primary materials employed in nonlinear dielectric nanophotonics. 

\begin{figure}[!h]\centering
\protect\includegraphics[width=0.41\columnwidth]{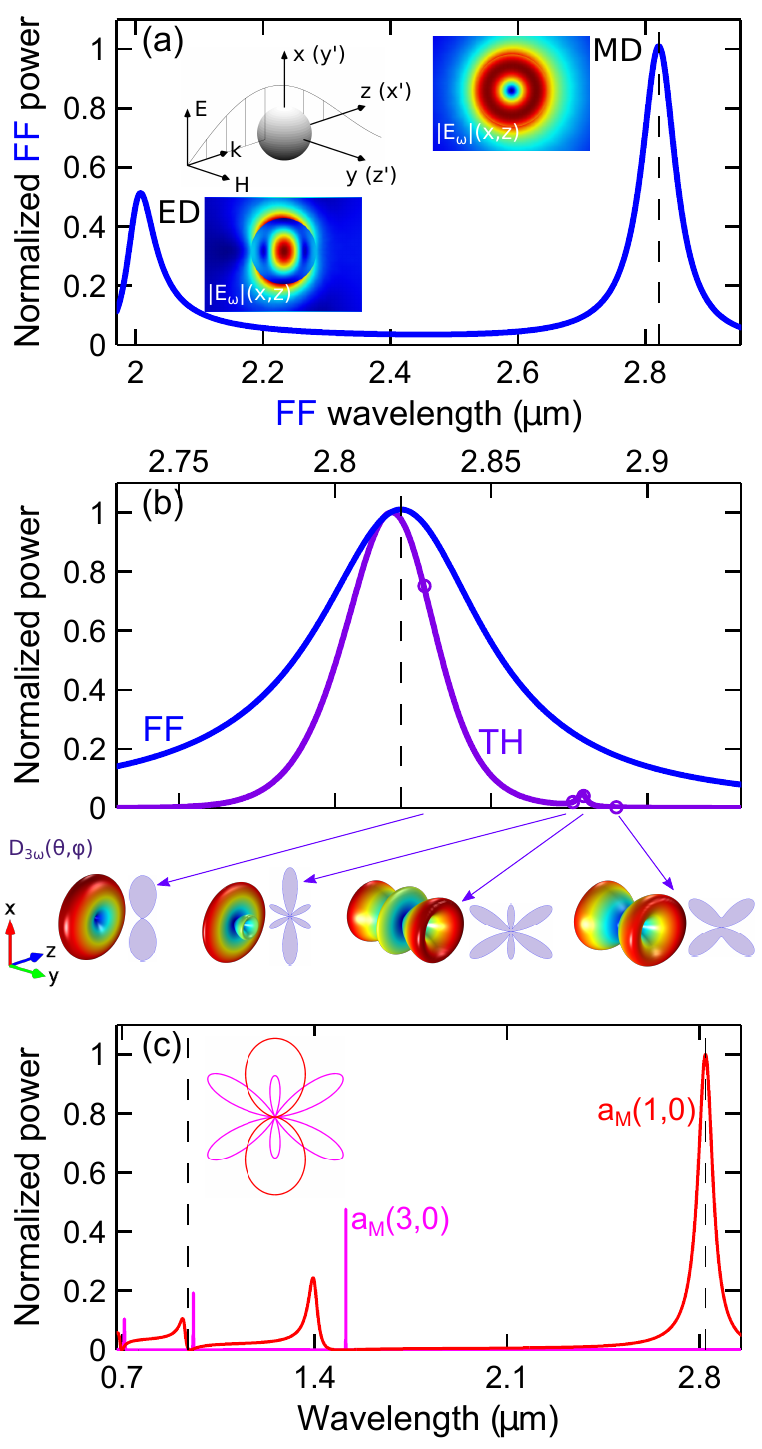}  \\
\caption{ {\bf Third-harmonic generation by a spherical high-index nanoparticle near the magnetic resonance.}
(a) Power spectrum of linear scattering calculated for a lossless high-index spherical dielectric particle of radius $a=230$ nm and refractive index $n = 6$, excited by the $x$-polarized plane wave. Insets: sketch of the geometry; the electric field norm distribution in the $xz$ plane at the electric and magnetic dipole resonances. (b) Scattered FF power (blue line) and radiated TH power (purple line) in the vicinity of the MD resonance as functions of the fundamental wavelength. Callouts illustrate TH radiation patterns and their cross-sections in the $xy$ plane at $4$ spectral points. (c) Magnetic dipole (red) and magnetic octupole (pink) contibutions to the linear scattering. Inset: cross-sections of radiation patterns for the magnetic dipole (red) and magnetic octupole (pink) in the $xy$ plane. Positions of the MD resonant wavelength and the respective TH wavelength are marked by vertical dashed black lines.
Powers are normalized to the magnetic dipole maximum values.
}
\label{fig:fig2}
\end{figure}

If the refractive index is high enough, electric (ED) and magnetic (MD) dipolar resonances are well-separated in frequency and rather narrow, as shown in Fig. S1(a). 
The analytically calculated TH power dependence, perfectly matching the numerical results obtained with the use of the finite-element method solver COMSOL Multiphysics, is plotted in Fig. S1(b) as a function of the wavelength of the incident radiation, where we also demonstrate variations of the TH emission pattern, originating from the interference of the magnetic dipole and magnetic octupole, though preserving azimuthal (with respect to the $y$ axis) symmetry. 
To elucidate a slight blue-shift of the THG peak from the fundamental MD wavelength, as well as appearance of the side small peak observed, in Fig. S1(c) we plot contributions from the magnetic dipole $a_M(1,0)$, $\propto 3 |a_{M}(1,0)|^2/k_0^2 $, and magnetic octupole $a_M(3,0)$, $\propto 7 |a_{M}(3,0)|^2/k_0^2 $, into the linear scattering over the wide range of wavelengths, based on the exact Mie solution. Each multipole has an infinite number of resonances. From our analysis, we expect only two multipoles to dominate others in the nonlinear response and visualize the spectral disposition of the generated modes around the tripled frequency. The small side peak is seen to be 
associated with the octupole resonance near the tripled frequency.

Despite the single mode approximation made in the derivation of the analytical theory, we notice that its applicability successively holds for the lower value of refractive index $n=3.5$ corresponding to silicon. The results of analytical and first-principles calculations in the proximity of the magnetic dipole resonance agree extremely well in both the cases $n=6$ and $n=3.5$ (the latter is presented in Fig. 2 of the main text). However, the magnetic dipolar and octupolar contributions to the nonlinear scattering are hidden 
under a common THG peak associated with the fundamental MD resonance for $n=3.5$, being clearly separated in frequency for $n=6$.

\newpage
\begin{figure}[t!]
\centerline{\mbox{\resizebox{8.9cm}{!}{\includegraphics{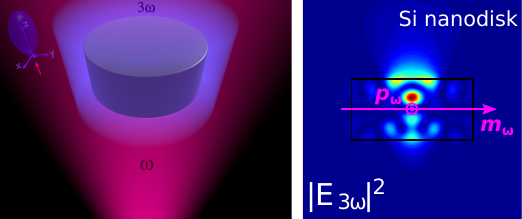}}}}
\end{figure}

\end{document}